%% file: main.tex
\documentclass{article}

\usepackage{microtype}
\usepackage{graphicx}
\usepackage{subcaption}
\usepackage{booktabs} 
\usepackage{xcolor}

\usepackage{hyperref}

\usepackage[preprint]{icml2026}

\usepackage{amsmath}
\usepackage{amssymb}
\usepackage{mathtools}
\usepackage{amsthm}

\usepackage[capitalize,noabbrev]{cleveref}

\theoremstyle{plain}

\theoremstyle{definition}

\theoremstyle{remark}

\usepackage[textsize=tiny]{todonotes}

\icmltitlerunning{The Next Paradigm Is User-Centric Agent, Not Platform-Centric Service}

\begin{document}

\twocolumn[
  \icmltitle{The Next Paradigm Is User-Centric Agent, \\ Not Platform-Centric Service}



  \icmlsetsymbol{equal}{*}

  \begin{icmlauthorlist}
    \icmlauthor{Luankang Zhang}{sch,equal}
    \icmlauthor{Hang Lv}{sch,equal}
    \icmlauthor{Qiushi Pan}{sch,equal}
    \icmlauthor{Kefen Wang}{sch,equal}
    \icmlauthor{Yonghao Huang}{sch,equal}
    \icmlauthor{Xinrui Miao}{sch,equal}
    \icmlauthor{Yin Xu}{sch}
    \icmlauthor{Wei Guo}{comp}
    \icmlauthor{Yong Liu}{comp}
    \icmlauthor{Hao Wang}{sch}
    \icmlauthor{Enhong Chen}{sch}
  \end{icmlauthorlist}

  \icmlaffiliation{comp}{Huawei Technologies, Shenzhen, China}
  \icmlaffiliation{sch}{University of Science and Technology  of China \& State Key Laboratory of  Cognitive Intelligence, Hefei, China}

  \icmlcorrespondingauthor{Hao Wang}{wanghao3@ustc.edu.cn}

  \icmlkeywords{Machine Learning, ICML}

  \vskip 0.3in
]



\printAffiliationsAndNotice{\icmlEqualContribution}  

\begin{abstract}
Modern digital services have evolved into indispensable tools, driving the present large-scale information systems. Yet, the prevailing platform-centric model, where services are optimized for platform-driven metrics such as engagement and conversion, often fails to align with users' true needs. While platform technologies have advanced significantly—especially with the integration of large language models (LLMs)—we argue that improvements in platform service quality do not necessarily translate to genuine user benefit. Instead, platform-centric services prioritize provider objectives over user welfare, resulting in conflicts against user interests. \textbf{This paper argues that the future of digital services should shift from a platform-centric to a user-centric agent.} These user-centric agents prioritize privacy, align with user-defined goals, and grant users control over their preferences and actions. With advancements in LLMs and on-device intelligence, the realization of this vision is now feasible. This paper explores the opportunities and challenges in transitioning to user-centric intelligence, presents a practical device-cloud pipeline for its implementation, and discusses the necessary governance and ecosystem structures for its adoption.

\end{abstract}

\vspace{-10pt}
\section{Introduction}
\input{Sections/1_Introduction}

\input{Sections/2_from_genrec_to_llmrec/2_from_genrec_to_llmrec}



\input{Sections/new_sec3/section3}

\input{Sections/4_Future_Application_Prospects/section_tmp}

\input{Sections/4_Future_Application_Prospects/section5}

\section{Alternative Views}
\label{sec:6}

\input{Sections/Alternative_views}

\section{Conclusion}
In conclusion, transitioning from platform-centric to user-centric agents represents a significant shift in digital service models. Platform-centric systems often prioritize provider objectives over user welfare, creating conflicts with user needs. This paper has highlighted the limitations of this approach and argued for a user-centric model that emphasizes user goals, privacy, and control. Advancements in LLMs and on-device intelligence make user-controlled agents feasible. We have proposed a device-cloud pipeline and discussed the governance structures needed to support this transition. Realizing user-centric intelligence will require both technological advancements and new governance frameworks to ensure privacy, security, and user autonomy.

\nocite{langley00}

\bibliography{example_paper}
\bibliographystyle{icml2026}




\end{document}

%% file: Sections/1_Introduction.tex
\begin{figure}[htbp]
\centering
\includegraphics[width=0.45\textwidth]{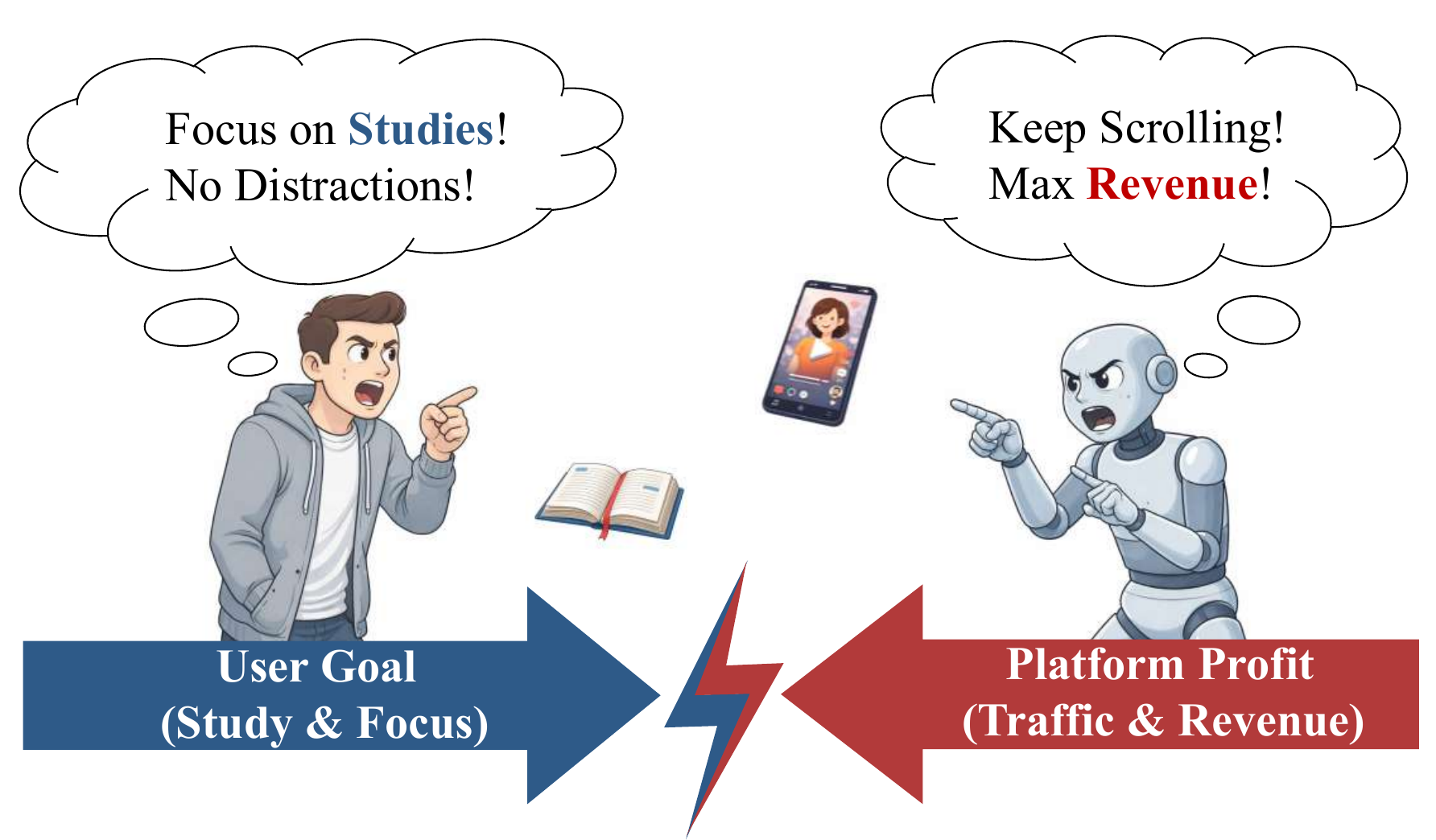}
\caption{Conflict of Interest: The platform-centric optimization objective fundamentally diverges from the user's goal.}
\vspace{-15pt}
\label{fig:1}
\end{figure}

Modern digital services have evolved into indispensable assistants, serving as the foundational infrastructure for today’s hyper-scale information ecosystems. Whether in content recommendations, business operations, or other service domains, providers typically observe user behavior within their platform, select candidate items from their own resources, and optimize for metrics such as engagement, retention, and conversion—all aimed at delivering personalized and efficient services. This \textbf{platform-centric} service paradigm remains the primary focus of both industrial deployment and academic research. The continuous evolution of technology, especially the application of Large Language Models (LLMs) across various service contexts\citep{achiam2023gpt}, has driven significant advancements in areas such as data\citep{luo2025survey}, model development\citep{zhou2025onerec}, and inference optimization\citep{zhang2024decodingaccelerationframework}, further enhancing service quality.

However, we raise a critical question: Can improvements in platform service quality truly translate into genuine user benefit? Our answer is no. We believe that service platforms optimized around platform goals cannot simultaneously act as trusted agents representing user interests. These services often prioritize platform objectives—such as advertising revenue, user retention, and conversion—over the user's needs. For instance, as shown in Figure~\ref{fig:1}, although a focused study might be the most beneficial for the user’s long-term development, platforms often push more video content to maximize ad revenue, a strategy that conflicts with the user’s broader goals. While platform-centric models have made impressive strides technologically, their core objectives often clash with users' true needs and interests.

Therefore, we argue that \textbf{the future service model should shift from platform-centric to user-centric intelligence}. In this model, user needs and interests take priority, with maximizing user utility being the primary goal, surpassing platform-driven profit motives. We propose that user-centric intelligence should be realized through a user-controlled agent, operating within the authorized personal context to execute and constrain goals and rules, while coordinating actions across platforms. This model offers several notable advantages compared to traditional platform services:
(1) \textbf{Privacy-by-design:} By adopting a “minimal disclosure/minimal necessary sharing” approach, sensitive personal contexts remain on the user side, rather than being uploaded to platforms.
(2) \textbf{Goal Alignment:} The agent’s decisions are driven by user-defined goals and constraints, not platform metrics like engagement or conversion.
(3) \textbf{User Agency:} Users maintain clear control and override capabilities over their preferences, constraints, and key actions, ensuring the service truly serves their needs.

\begin{figure}[t]
\centering
\includegraphics[width=0.48\textwidth]{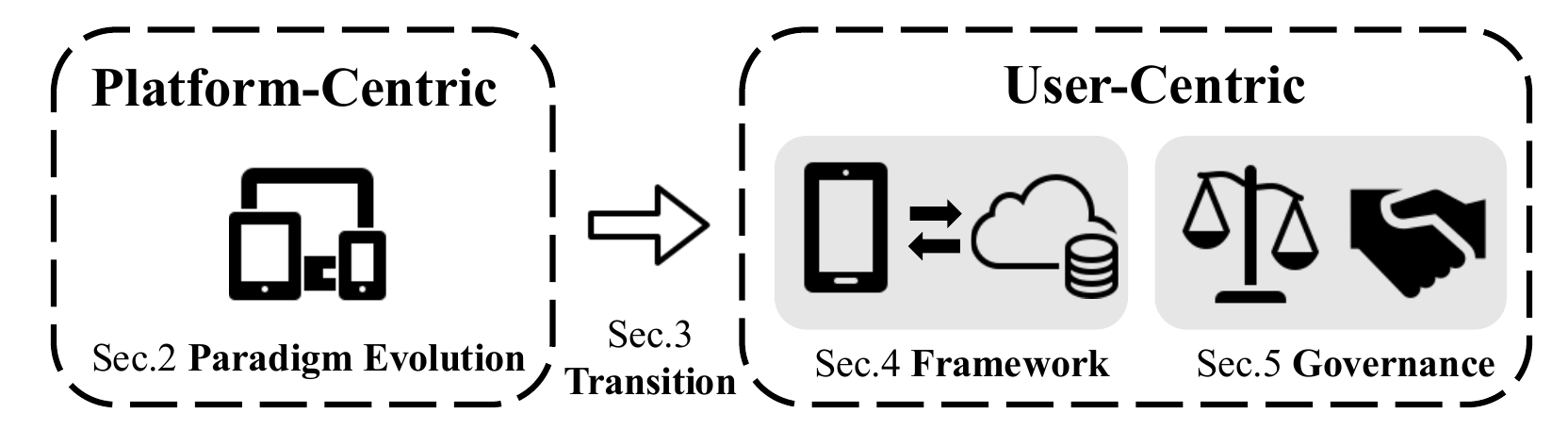}
\vspace{-15pt}
\caption{Organization of Our Paper}
\label{fig:2}
\vspace{-20pt}
\end{figure}

With the maturity of technologies such as LLM-based intent reasoning and on-device intelligence, achieving such a user-centric intelligence is no longer an unattainable goal. This paper explores the opportunities and challenges in realizing this vision, structured as follows:

\textbf{Roadmap and Structure:} As illustrated in Figure~\ref{fig:2}, we begin by summarizing the technological evolution driving the shift from platform-centric to user-centric services, focusing on the current state of platform-centric services and their inherent limitations (Section~\ref{sec:2}). We then explain why user-centric intelligence can effectively address the three major structural bottlenecks in current platform services: fragmented context, limited execution boundaries, and misaligned incentives (Section~\ref{sec:3}). Next, we propose a practical “device-cloud” pipeline model to implement user-centric intelligence (Section~\ref{sec:4}). Finally, we discuss the governance structures and ecosystem contracts needed to promote such intelligence (Section~\ref{sec:governance}). Additionally, we compare alternative perspectives and further discuss various existing challenges and future ideas (Section~\ref{sec:6}).

%% file: Sections/2_from_genrec_to_llmrec/2_from_genrec_to_llmrec.tex
\section{Evolution of Platform-Centric Service}
\label{sec:2}

\input{Sections/2_from_genrec_to_llmrec/2_0_Overview}

\subsection{Optimization of Data Pipeline}
\label{sec2:data}
\input{Sections/2_from_genrec_to_llmrec/2_1_Data}

\subsection{Advancement of Model Frameworks}
\label{sec2:model}
\input{Sections/2_from_genrec_to_llmrec/2_2_Model}

\subsection{Refinement of Inference Strategies}
\label{sec2:infer}
\input{Sections/2_from_genrec_to_llmrec/2_3_Inference}

%% file: Sections/2_from_genrec_to_llmrec/2_0_Overview.tex

Most large-scale information services today operate as platform-centric services following an inner loop: the platform observes what users do inside one app, decides what to show next from its own inventory, and learns from the resulting feedback to improve future decisions. Data, objectives, and guardrails for these services are all determined by the provider, including which signals are collected, which outcomes are optimized (often engagement or conversion), and what can or cannot be shown under various policy, product, and latency constraints. Recent years have brought major technical upgrades, from early generative pipelines that treat user histories as sequences of platform-specific IDs to foundation models that can understand text and produce natural-language outputs. 
In the following sections, we provide a detailed analysis of this platform-centric service paradigm across three fundamental dimensions: the optimization of data pipelines (Section~\ref{sec2:data}), the advancement of model frameworks (Section~\ref{sec2:model}), and the refinement of inference strategies (Section~\ref{sec2:infer}).


%% file: Sections/2_from_genrec_to_llmrec/2_1_Data.tex
\begin{figure*}
    \centering
    \includegraphics[width=0.99\linewidth]{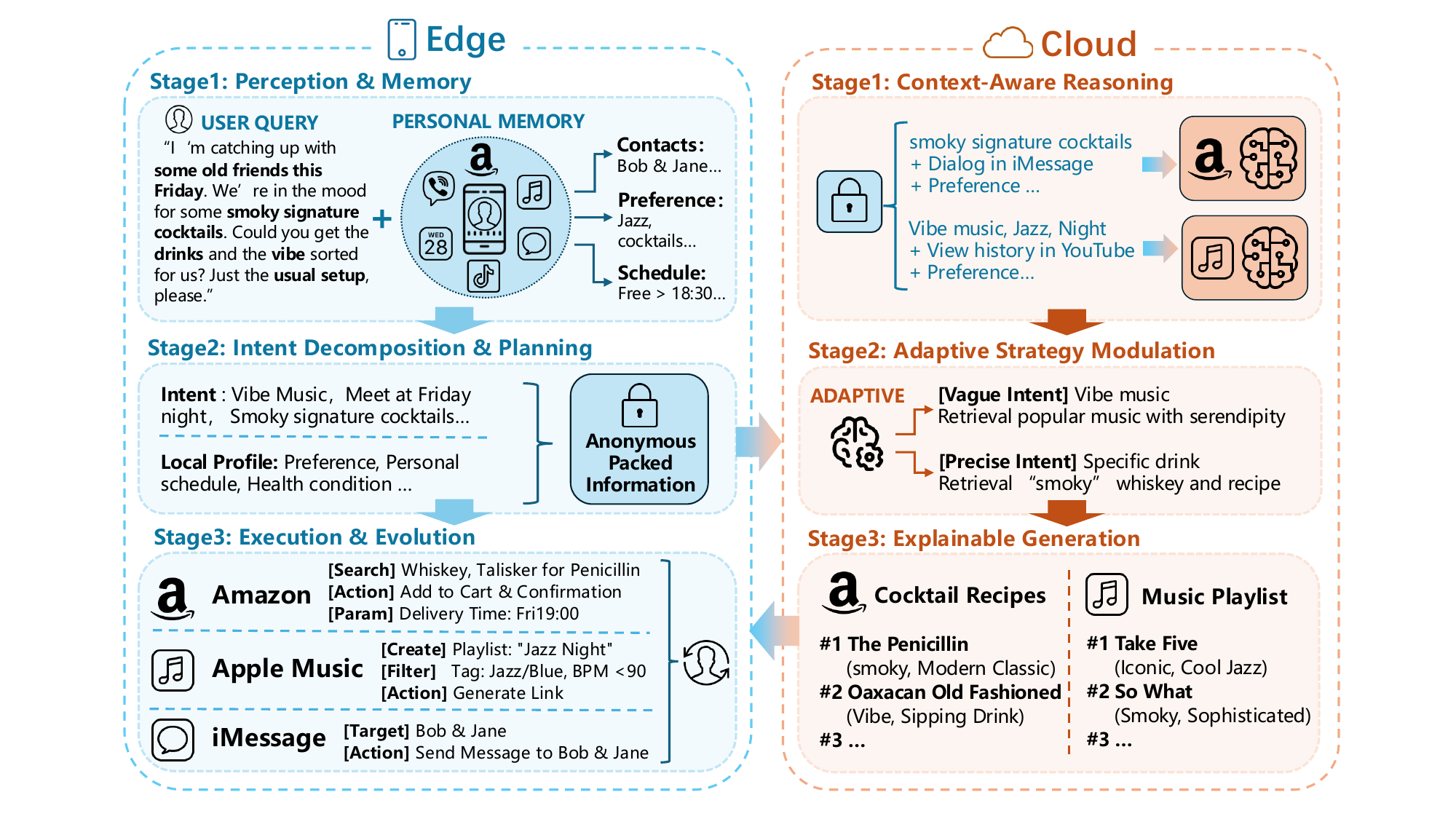}
    \vspace{-10pt}
    \caption{An Edge-Cloud Collaborative Pipeline for User-Centric Agent}
    \label{fig:placeholder}
    \vspace{-15pt}
\end{figure*}

Prior data-centric research in platform-centric service can be summarized into three pipeline stages: evaluation, generation, and compression. \textbf{(1) Data Evaluation} aims to quantify data quality and relate data/model scale to downstream performance, often through scaling laws \citep{hoffmann2022training}. In practice, many pipelines rely on static, model-driven proxies (e.g., entropy- or perplexity-based heuristics) to effectively characterize samples \citep{soldaini2024dolma, penedo2024fineweb}. Recent work further adopts dynamic attribution methods—such as Influence Functions or Shapley values—to precisely estimate the marginal utility of data subsets during the training process \citep{li2024shapley}. \textbf{(2) Data Generation} targets data sparsity by synthesizing additional training signals. Representative approaches include distribution matching \citep{liu2023diffusion}, heuristic augmentation based on statistical regularities \citep{yin2024dataset, lee2025sequential}, and knowledge-augmented generation that injects external information beyond platform data \citep{lyu2024llm}. \textbf{(3) Data Compression} reduces training cost under redundancy. Data selection filters less informative samples using various scoring criteria \citep{chen2023alpagasus, liu2023makes}, while data distillation constructs compact summaries that preserve performance with fewer training steps \citep{zhang2025td3, yin2023squeeze}.


However, these improvements rely on platform-specific discrete IDs that define item identity only within a single platform. This makes semantics implicit and hinders reuse, exacerbating cold-start and cross-domain challenges \citep{panda2022approaches, zhu2021cross}. While remedies like hierarchical indexing or side information fusion exist, representations remain platform-bounded and optimized within isolated collaborative spaces. This motivates the LLM-based shift: by adopting universal semantic tokens, this paradigm transitions item modeling from closed collaborative spaces to an open, transferable semantic space.


The transition to LLM-based paradigm changes data requirements. Beyond modeling behavioral co-occurrence, training now needs supervision that supports semantic grounding and explicit reasoning. As a result, data-centric service has concentrated on two aspects, namely data scarcity and data mixture.  (1) \textbf{Data Scarcity} stems from a structural lack of high-quality alignment signals that connect implicit behaviors to explicit intent and reasoning. One route is self-evolving generation, where the model iteratively synthesizes and refines training examples using its own generative capability \citep{yuan2024self, madaan2023self}. Another route is reasoning-aware synthesis, which converts interaction traces into explicit Chain-of-Thought supervision through LLM annotation or human labeling, so that decision logic becomes directly learnable \citep{zhou2025openonerec, yue2025cot4rec}.  (2) \textbf{Data Mixture} addresses the optimization tension between preserving general capabilities and adapting to specific tasks. Strategic sampling mitigates catastrophic forgetting by tuning co-training ratios between specific domain data and general corpora \citep{liao2024llara, ye2024data}. In addition, conflict-aware mixture strategies adjust domain mixture weights using domain-level training signals, such as loss-driven reweighting under a minimax objective, so that domain adaptation does not degrade core instruction-following ability \citep{xie2023doremi}.

%% file: Sections/2_from_genrec_to_llmrec/2_2_Model.tex
Recent advances in platform-centric models are mainly trained from scratch, which contain three aspects: item tokenization, efficient architectures, and multi-objective alignment.  
(1) \textbf{Item Tokenization} emerges as a solution to the storage constraints caused by the massive embedding tables required for traditional random item IDs. This paradigm leverages item metadata to encode items into sequences of discrete tokens \citep{rajput2023recommender}. Some works attempt to integrate text information and collaborative information during the tokenization process \citep{wang2024learnable,deng2025onerec,xiao2025progressive}. Additionally, to achieve more expressive item representations, some efforts focus on advanced tokenizers, such as parallel or context-aware tokenizers \citep{hou2025generating, hou2025actionpiece, zhong2025pctx}.
(2) \textbf{Efficient Architectures} are developed to tackle the computational overhead associated with standard self-attention mechanism, particularly when modeling long user interaction sequences on large-scale datasets. Some researches focus on designing efficient Transformer variants to reduce computational costs \citep{zhai2024actions,ye2025fuxi,zhou2025onerec, ye2025fuxi, zhou2025multi,guo2024scaling,pan2025revisiting,xie2024breaking,zhou2026survey}, while others are exploring the replacement of the Transformer backbone entirely with State-Space Models (SSMs) to achieve better computational efficiency \citep{liu2024mamba4rec,liu2025sigma}. 
(3) \textbf{Multi-Objective Alignment} aims to bridge the traditional processing stages and broaden the scope of optimization beyond simple accuracy. Some efforts unify distinct phases such as retrieval and ranking into a generative framework \citep{deng2025onerec,zhang2025killing}. Another type of works expand the optimization by incorporating auxiliary objectives, such as novelty and diversity to ensure that the model satisfies a more comprehensive range of user needs \citep{meng2025generative,xu2025multi,wang2025mf,zhang2024unified}.

However, models trained from scratch face a fundamental issue: they tend to fit specific data distributions rather than achieving comprehension of the rich semantic information. Even when text-based tokenization is employed, these models just enhance the item representations based on text similarities. Unlike LLMs which possess broad knowledge and reasoning capacity derived from large-scale data, traditional models can only rely on specific datasets. Consequently, the transition to LLM-based services is essential to address these limitations, enabling the platform-centric service to directly leverage the generalized understanding and reasoning capability inherent in LLMs.

The advancement of the LLM-based paradigm has been primarily driven by two critical lines of research aimed at adapting LLMs to information service scenarios \citep{wang2025generative}, (1) \textbf{Modality Alignment}, which focuses on resolving the discrepancy between user preference prediction and natural language understanding. Early approaches circumvented this issue by formulating user preference prediction as a pure text generation task \citep{raffel2020exploring,cui2022m6,sanner2023large,ji2024genrec}, directly combining item descriptions as input, while more recent methodologies have evolved to utilize discrete item tokenizers as a semantic bridge, effectively connecting collaborative signals with the textual modality through learnable tokens representing items \citep{zheng2024adapting,hong2025eager,wang2025empowering,liu2025onerec,he2025plum}.
(2) \textbf{Reasoning Capability} refers to enabling the model to deeply understand the user's intentions during the information service process. The prevailing methods mainly involve activating the reasoning capability by fine-tuning on the data containing chain-of-thought, as well as reasoning enhancement through RL-based post-training\citep{liu2025onerec,kong2025think}.

%% file: Sections/2_from_genrec_to_llmrec/2_3_Inference.tex
The existing platform-centric generative inference paradigm defines the inference process as a sequence completion task based on in-app data. To extract complex intents from platform-specific interactions and map them into executable actions, this paradigm primarily relies on \textbf{(1) autoregressive sequence generation}, which leverages Transformer architectures to encode user behaviors into latent representations~\cite{rajput2023recommender, wang2024learnable} and refines next-item predictions through greedy search, beam search, and test-time scaling~\cite{muennighoff2025s1} techniques. To ensure the validity of generated identifiers within the platform's catalog, \textbf{(2) search space pruning paradigms} lock valid branches by performing step-wise restrictions on structures like prefix trees~\cite{penha2025semantic, wu2025constrained, liao2025avoid}, or employ graph constraints and quantization optimization for multi-path concurrent verification~\cite{hou2025generating, wang2025nezha} to ensure ID legitimacy. To address the inference latency bottleneck inherent in autoregressive decoding, \textbf{(3) non-autoregressive mechanisms and iterative distribution modeling} have emerged as an alternative core paradigm; its technical route includes multi-token synchronous prediction~\cite{gloeckle2024better, wang2025act} and the use of discrete diffusion models~\cite{gao2025mindrec, shah2025masked, liu2025diffgrm} and flow matching networks~\cite{liu2025flow_cf, li2025preference, liu2025flow_sr} for accelerated processing of platform-side structured data.

However, these methods fall short in complex service fulfillment compared to LLMs. By treating inference as passive sequence completion, the models' reliance on data fitting limits them to mimicking historical distributions, obscuring the deep semantics and world knowledge required for sophisticated needs~\cite{vats2024exploring, li2024large}. Structural rigidity also restricts instruction following~\cite{hao2025oxygenrec} and dynamic interaction~\cite{huang2025survey} due to inputs being confined to local history. Consequently, generation often lacks logical consistency and explainability~\cite{ma2024xrec, zhang2025oracle}, favoring opaque probabilities over explicit reasoning. These constraints necessitate a transition from implicit matching toward paradigms centered on open-ended instruction following and explicit logic.

Integrating LLMs enhances reasoning but conflicts with industrial latency constraints. To bridge this gap, holistic strategies first reduce global workload via model distillation~\cite{xu2024slmrec}, mixed-precision quantization~\cite{huang2024billm}, or cascade routing~\cite{chen2023frugalgpt}, further dissecting optimization into two physical bottlenecks: \textbf{(1) Prefill Optimization: Breaking the Context Bottleneck.} To optimize the first-token latency, semantic compression reduces input length by pruning redundant contexts~\cite{jiang2023llmlingua} or aggregating behavior segments~\cite{chai2025longer, liu2024mamba4rec}, while dynamic sparsity bypasses quadratic complexity through progressive token dropping~\cite{fu2024lazyllm}, layer-wise sparsity inversion~\cite{yang2025earn}, or pattern-based sparse masks~\cite{jiang2024minference} to achieve linear complexity. \textbf{(2) Decoder Optimization: Accelerating Serial Generation.} To accelerate bandwidth-constrained serial generation, speculative and parallel decoding verify multiple tokens via draft models~\cite{li2024eagle} or non-autoregressive paradigms~\cite{lin2024efficient}, while resource efficiency is maximized through adaptive computation~\cite{kumar2025helios} and KV cache compression~\cite{ge2023model} to reduce memory footprint without compromising service quality.

%% file: Sections/new_sec3/section3.tex

\section{Toward User-Centric Agents}
\label{sec:3}

Service systems have made remarkable technical progress, from collaborative filtering to LLM-based reasoning, yet the core paradigm has barely changed~\cite{huang2025survey}. They remain fundamentally \textbf{platform-centric}: trained on proprietary data locked inside walled gardens, and optimized for platform-defined objectives such as ad revenue, retention, and lock-in. Put simply, today’s services are not designed to serve users first; they are designed to serve platforms by maximizing platform utility within closed loops.

A platform-centric model cannot simultaneously function as a user’s trusted agent and a system optimized for platform profit. A more capable model is not a more benevolent model. It is a stronger amplifier of its objective. When the reward is retention and revenue~\cite{deng2025onerec, zhang2025gpr}, better models become increasingly proficient at reinforcing the ``walled garden" effect and monetizing user attention, rather than maximizing true user utility.


To break this deadlock, we advocate a structural inversion toward a \textbf{User-Centric Paradigm}. This paradigm shifts the locus of control from platforms to users, anchoring intelligence in a personal AI agent that is fully owned and controlled by the user. Freed from platform objectives, this agent can act as the user’s fiduciary. It aggregates on-device signals to form a more holistic context, orchestrates cross-service workflows to fulfill intents, and aligns actions with user utility rather than platform profit. In short, the platform-centric model does not merely fail to deliver these capabilities; it systematically suppresses them.

\subsection{From Fragmented Logs to Holistic Context}
\label{sec:context}

Reliable assistance requires a holistic context. What a user should see, when they should see it, and whether it is even appropriate to interrupt them depends on signals that lie far beyond in-app clicks. The same service can be helpful at one moment and harmful at another if the system is blind to the user’s state, such as current screen activity, recent cross-app actions, time sensitivity, and private constraints like meetings, location, and financial budget.

Yet under the platform-centric regime, context is structurally fragmented. Each platform only learns what happens inside its own product surface, while the signals that define the user’s real situation are scattered across apps and device-level traces. Crucially, this gap cannot be closed within the platform-centric paradigm. Platforms do not and cannot observe the full cross-app and private signals needed for a holistic context. Pushing such context into provider-controlled clouds is untenable at scale, both economically and in terms of privacy~\cite{zhang2024enabling, zhang2025personalized}. As a result, platform-centric systems are structurally confined to fragmented logs, and their service is built on an incomplete and biased view of the user.

A user-centric agent resolves this constraint by inverting the data flow~\cite{xu2025iagent}. Holistic context already converges on the device, where the user’s activities naturally span applications and modalities. When intelligence is anchored locally and governed by the user, the agent can integrate private signals without exporting them by default. This shifts personalization from a data-extraction problem into a local reasoning problem, enabling rich context awareness while keeping sensitive information under user control.

\subsection{From Item Prediction to Intent Fulfillment}
\label{sec:intent}

Platform-centric service was built for a narrower objective: predicting the next item a user might click within a fixed candidate pool. Its success comes from optimizing engagement primitives such as CTR and dwell time, and from keeping the user inside a single application loop~\cite{zhou2025onerec}. This design is effective for simple passive consumption, but it inevitably fails when the user’s goal is not merely to browse but effectively to get something done.

In the agentic era, users increasingly delegate intents that require planning and action across services~\cite{zhang2025api}. These intents are inherently procedural, for example, rebooking a missed flight, negotiating a refund, coordinating a multi-party meeting, or switching a subscription plan under constraints. A platform-centric system is structurally unfit for such requests. It is confined to a single platform boundary and governed by an objective that rewards retention over resolution. Even with LLM-level reasoning, a platform-centric system can only execute what lies within its own service boundary. It may complete in-app actions, but it cannot reliably orchestrate end-to-end workflows that span multiple platforms, tools, and external commitments.

A user-centric agent extends execution beyond any single platform boundary. It treats an intent as an end-to-end plan to be completed under user constraints, rather than a query answered by an in-app ranked list. With holistic on-device context and cross-service tool use, the agent can decompose goals into steps, coordinate information and actions across providers, and carry the workflow to completion. This shifts the metric of success from engagement to utility, and from ``what to show next'' to ``what to do next''.


\subsection{From Platform Objectives to User Utility}
\label{sec:incentives}


In the agentic era, the real power lies not only in what the system recommends, but in what it \emph{decides} and \emph{executes} on behalf of the user. As user-centric agents gain the ability to take actions, rather than merely suggesting options, the most impactful decisions become those that are often invisible to the user: which options are considered, which alternatives are excluded, which provider is selected by default, and how conflicts between constraints are resolved. These decisions, made autonomously by the agent, ultimately shape the outcomes of the user’s journey.

A platform-centric system places these crucial decisions under the direct control of the platform itself. Platform incentives become apparent in these decision layers, which are often subtle and indirect. While steering users toward specific actions may seem harmless, it does not always require blatantly bad advice. Instead, it can occur through small, seemingly reasonable choices that favor the platform’s own services. For example, platforms may prioritize their own products by default, create friction for competitors, or select the most profitable default actions to maximize revenue~\cite{dash2021umpire}. Over the course of a multi-step workflow, these small biases accumulate, systematically directing the outcome toward platform interests—even when each individual step appears to benefit the user.

This is why trust cannot be restored by simply making models smarter. Increased capability only makes the system better at optimizing its own reward, which, in a platform-centric model, is not aligned with user utility. The only lasting solution is governance. A user-centric agent makes this separation explicit: the user defines objectives and constraints, and the agent enforces them when selecting options and executing actions. Platforms may still provide inventory and services, but they no longer control the decisions made on behalf of the user. User utility becomes the primary objective, not a secondary consideration.



\noindent\textbf{Why now.}
Historically, the transition to a user-centric architecture was constrained by engineering realities~\cite{gill2024modern}. A personal agent requires both a deep understanding of user intent and the ability to execute securely under user control. We have now reached a critical inflection point on both fronts. First, the paradigm of LLM-as-Rec upgrades recommendation from ID-based matching to intent-level reasoning over semantic context~\cite{liu2025onerec}. Second, on-device intelligence has advanced to the point where a high-performance compact LLM can run directly on user hardware~\cite{liu2024mobilellm}. With understanding and execution finally co-located on the device, user-centric agents move from a theoretical ideal to an engineering reality.

%% file: Sections/4_Future_Application_Prospects/section_tmp.tex
\section{Realizing User-Centric Agents}
\label{sec:4}

The structural constraints outlined in \S \ref{sec:3} require more than a conceptual update; they demand a comprehensive and concrete architectural shift that can be implemented as an executable pipeline. A purely on-device solution is insufficient. While the device is the only place where holistic personal context can be governed by the user, many intents require direct interactions with the vast external service ecosystem—real-time inventories, rapidly changing prices, and authenticated endpoints for booking, payment, and cancellation. The device can reason locally, but it cannot manage the world’s inventories or control service endpoints.

To address this, we propose an edge-cloud collaborative pipeline centered on an on-device personal agent. The device-side agent serves as the user-controlled interface: it synthesizes local context to interpret intent, enforces private constraints, and makes final decisions. The cloud acts as a callable service layer, providing access to external inventories, tools, and execution interfaces. In this model, the cloud is no longer a gatekeeper dictating outcomes; instead, it responds to structured requests from the device and returns candidates with interpretable rationales, allowing the device to select actions that maximize user utility.



\subsection{Constructing the On-Device Agent}
\label{sec:4.1}

Building a sovereign agent requires more than porting a language model to mobile hardware; it demands a cohesive cognitive system capable of translating private, local signals into verifiable actions. The on-device architecture is framed as a closed-loop lifecycle comprising three stages: Perception, Planning, and Execution.

The lifecycle begins with \textbf{Perception and Memory}. Unlike cloud systems that rely on fragmented platform logs, an on-device agent can access user-centric signals through OS-level instrumentation, such as cross-app activity, notifications, calendar events, and device context (e.g., time and location). The challenge here is not access, but \emph{context construction}. These signals are heterogeneous and time-dependent, and only a small subset is relevant to any given task. Therefore, the agent must synthesize raw events into a compact \emph{situational state} that captures the user's current actions, active constraints, and missing information—providing a stable foundation for downstream planning \cite{contextagent,contextllm}. Memory then makes this state reusable without overwhelming the device’s limited resources. Rather than being a static archive, memory is tiered and evolving: short-term context for the current session, episodic traces of recent outcomes, and long-term semantic preferences that update gradually \cite{memgpt}. Beyond retrieval, the key operation is \emph{consolidation}, where repeated experiences are compressed into durable summaries or rules that improve future decision-making \cite{memagent2,memagent1}.

The next stage is \textbf{Intent Decomposition and Planning}. The agent interprets abstract goals by grounding them in the memory profile and current device state, then decomposes the request into a sequence of decisions rather than a single-shot response. The core challenge is deliberation: real-world intents are often underspecified, constraint-heavy, and multi-step, so the agent must propose multiple candidate plans, evaluate feasibility under constraints, and adjust when early choices lead to dead ends. This approach aligns with deliberative planning frameworks that explicitly support exploration of alternative reasoning trajectories and plan structures \cite{treesearch,treesearch2,plangen}. Planning also determines what information must be gathered and when. Instead of entangling reasoning with every intermediate observation, the agent drafts a structured plan with required evidence and checkpoints \citep{gu2025rapid}, filling in missing pieces through targeted queries—enhancing robustness in scenarios of partial observability and tool uncertainty \cite{rewoo}. In a user-centric setting, planning further defines delegation boundaries: it decides what to outsource to the cloud (e.g., public inventories and execution endpoints) and what to resolve locally (e.g., private constraints and user policies). If an intent remains ambiguous, the plan includes clarification steps, either as concise questions to the user or structured follow-up requests to services.

The cycle concludes with \textbf{Execution and Evolution}. The agent queries external services through the cloud and receives candidates from the public service space, but the final decision is made on-device. The key role of this stage is to ensure that execution is verifiable under user control. Before committing to any irreversible action, the agent performs a lightweight runtime check, filtering candidates against private constraints and user policies (e.g., calendar feasibility, budget limits, and explicit confirmation for sensitive actions). It can also locally re-rank results using private preferences that should not be exported. This action-level constraint enforcement is increasingly studied in agent safety and runtime enforcement frameworks, which evaluate proposed actions against user-defined rules before execution \cite{agentspec,guardagent,agrail}. After execution, outcomes and user feedback are written back into memory to refine what is retrieved and how constraints are prioritized in future plans, forming a self-improving loop that strengthens personalization through local consolidation rather than centralized accumulation of raw private context \cite{memagent2,memagent1}.

\subsection{The Cloud Evolution}
\label{sec:4.2}

To support the on-device agent's capabilities, the cloud must evolve from a passive ranking endpoint to an agentic service layer. The shift is functional: instead of optimizing engagement in a closed loop, the cloud assists the on-device agent by solving external problems at scale—such as querying dynamic inventories, applying public constraints, and providing verifiable candidates. In this model, the cloud responds to structured requests encoding intent and constraints, while raw signals remain controlled by the device.


The loop begins with \textbf{Context-Aware Reasoning}. The cloud agent ingests compact context from the device, including task goals, preferences, and explicit constraints, to reason over the external service space. A key challenge arises from the fact that real-world requests are often underspecified or evolve over time, necessitating that cloud reasoning maintains a coherent multi-turn state. This allows the cloud to generate clarification questions and refine its responses, rather than providing a one-time list of options \cite{taira,context1,context2,context3}. Conceptually, the cloud shifts from ranking items to participating in an interactive decision process that is driven and governed by the user-side agent.

Next, the loop progresses with \textbf{Adaptive Strategy Modulation}. User intents span various modes, and the same retrieval or ranking strategy is not suitable for all of them. For time-sensitive requests, the cloud prioritizes constraint satisfaction and precision; for exploratory tasks, it expands the search space and supports controlled novelty and serendipity. This strategy adapts to the user’s current task state, modulated by signals from the device, aligning cloud behavior with the user’s objectives rather than defaulting to engagement-maximizing heuristics \cite{strategy3,strategy4,strategy1,strategy2}.

The loop concludes with \textbf{Explainable Generation}. In an agent-mediated pipeline, opaque results are insufficient because the final decision is made and verified on-device. The cloud, therefore, provides interpretable rationales and machine-verifiable metadata for each candidate, explicitly linking options to received constraints and surfacing key trade-offs \cite{reason1,reason2,reason3}. This output must be structured, so the on-device agent can reliably parse constraint tags, evidence, and justifications for local validation \cite{structure1,structure2}. In this way, transparency becomes a core feature of the device-cloud pipeline, rather than an afterthought.

%% file: Sections/4_Future_Application_Prospects/section5.tex
\section{Governing User-Centric Agents}
\label{sec:governance}

The edge--cloud pipeline described above answers \emph{how} a user-centric agent could operate, but it does not guarantee that it will be adopted. The harder barrier is institutional. A user-centric agent reshapes control over two assets that define the modern attention economy: (i) the \emph{interface} where user intent is expressed and routed, and (ii) the \emph{execution path} where transactions are completed. Once an agent mediates these two steps, it becomes the de facto traffic allocator and decision broker. This creates a governance and bargaining problem that is orthogonal to model quality: \emph{who controls the agent, and how do platforms and agents share value without recreating platform-centric capture?}

\subsection{Who Provides the Agent? }
\label{sec:agent-provider}

Control over the on-device agent is not a neutral implementation detail; it is an economic vantage point. The entity that ships the agent can shape defaults, determine which services are discoverable, and influence the terms under which third-party platforms are invoked. In other words, replacing ``apps as the interface'' with ``agent as the interface'' does not automatically yield user sovereignty. If the agent itself becomes a proprietary gatekeeper, the ecosystem risks simply swapping one platform-centric bottleneck for another.

This tension suggests that the central question is governance rather than capability. A viable user-centric ecosystem requires that the agent be \emph{user-governed} in enforceable ways: users must be able to inspect and override preferences, constrain sensitive actions, and switch providers without losing their personal state \cite{lazar2025moral,kapoor2025build}. This implies portability of personal memory, transparent policy surfaces, and a separation between the user-controlled decision layer and any single vendor’s commercial objectives. Without these safeguards, the agent provider inherits the same incentive to extract rents from attention and transactions, undermining the motivation for a user-centric shift.

\subsection{How Do Agents and Platforms Coordinate? }
\label{sec:agent-platform}

Even with a user-governed agent, the ecosystem cannot function without deep cooperation from platforms, because platforms own the inventories and execution endpoints that make intents actionable. Yet this cooperation is economically non-trivial. In the app-centric regime, platforms defend the direct user interface because it enables retention, upsell, and branding. An agent that shortcuts the user journey threatens to commoditize platforms into interchangeable backends, weakening their ability to capture value.

A sustainable equilibrium therefore requires a \emph{value exchange} rather than moral persuasion. The agent needs execution access; the platform needs qualified intent and high-conversion matching. This motivates a negotiated contract that can be summarized as a trade of \textit{Authority for Intent}: platforms expose standardized, intent-level APIs and reliable execution interfaces, while the agent provides structured intent representations, constraint summaries, and attribution signals that improve conversion and reduce waste. Crucially, this exchange must be designed to preserve user sovereignty. The agent should transmit the \emph{minimum sufficient} context for the requested action, with user-controlled consent boundaries, while keeping raw private context local.

This coordination cannot rely on ad hoc integration alone. It requires new shared primitives at the ecosystem level: (i) \emph{standardized intent and constraint schemas} so platforms can respond to agent requests without owning the full user profile; (ii) \emph{auditability and verifiable compliance} so both sides can prove what constraints were requested and satisfied; and (iii) \emph{economic settlement mechanisms} for attribution and revenue sharing, otherwise platforms have no incentive to open deep execution interfaces. In effect, user-centric agents turn recommendations into a two-sided market with a new control plane. Making that market competitive and trustworthy demands explicit rules, not only better models.

\noindent\textbf{Putting it together.}
The user-centric paradigm is therefore essentially a fundamental socio-technical transition. Engineering makes the pipeline feasible, but governance determines whether the agent becomes a user fiduciary or the next gatekeeper. The long-run success of user-centric agents hinges on designing interoperability contracts and market mechanisms that align transaction volume and user utility, while preventing control over the intent interface from collapsing back into platform-centric capture.

%% file: Sections/Alternative_views.tex
Finally, we examine two paradigms currently being explored: \textbf{System-level Agents}, which possess rich context but lack capability, and \textbf{Service-level Agents}, which possess capability but lack context.

\subsection{The System-level Context-driven Paradigm}
This paradigm, represented by assistants like Apple Intelligence or Doubao, leverages \textbf{system-level privileges} to capture broad ambient context. Some entrants employ vision-based UI automation to parse screens and simulate interactions, theoretically enabling unified app scheduling without explicit vendor authorization.

\textbf{Our Rebuttal:} Despite their ``God's eye view," these agents lack execution depth. They remain confined to the OS surface, unable to access complex internal business logic within third-party applications. Furthermore, reliance on fragile visual simulation without native API support introduces instability and non-determinism, making them unsuitable for high-reliability tasks such as finance or security.

\subsection{The Service-level Execution-driven Paradigm
}
This paradigm emphasizes \textbf{professional execution}, led by vertical specialists such as Meituan and Didi or aggregators such as Qwen utilizing plugins. Supporters argue these platforms provide the most reliable entry points due to their established service infrastructure and payment loops.

\textbf{Our Rebuttal:} This paradigm suffers from contextual isolation. Operating strictly at the application layer, these agents lack access to underlying OS data and cross-app behavior trajectories. This information deficit forces reactive responses rather than proactive intent fulfillment, severely limiting their utility in complex, multi-stage scenarios